\documentclass[nofootinbib,twocolumn,aps,letterpaper,superscriptaddress,showpacs]{revtex4}
\usepackage{amsmath}
\usepackage{amsfonts}
\usepackage[dvips]{graphicx}
\begin{document}

\newcommand\be{\begin{equation}}
\newcommand\ee{\end{equation}}
\newcommand\bea{\begin{eqnarray}}
\newcommand\eea{\end{eqnarray}}
\newcommand\bseq{\begin{subequations}} 
\newcommand\eseq{\end{subequations}}
\newcommand\bcas{\begin{cases}}
\newcommand\ecas{\end{cases}}
\newcommand{\p}{\partial}
\newcommand{\f}{\frac}

\title{Scalar Field Theory on Non-commutative Snyder Space-Time}

\author{Marco Valerio Battisti}
\email{battisti@icra.it}
\affiliation{Centre de Physique Th\'eorique, Case 907 Luminy, 13288 Marseille, France}
\author{Stjepan Meljanac}
\email{meljanac@irb.hr}
\affiliation{Rudjer Boskovic Institute, Bijenicka c.54, HR-10002 Zagreb, Croatia}


\begin{abstract}
We construct a scalar field theory on the Snyder non-commutative space-time. The symmetry underlying the Snyder geometry is deformed at the co-algebraic level only, while its Poincar\'e algebra is undeformed. The Lorentz sector is undeformed at both algebraic and co-algebraic level, but the co-product for momenta (defining the star-product) is non-co-associative. The Snyder-deformed Poincar\'e group is described by a non-co-associative Hopf algebra. The definition of the interacting theory in terms of a non-associative star-product is thus questionable. We avoid the non-associativity by the use of a space-time picture based on the concept of realization of a non-commutative geometry. The two main results we obtain are: (i) the generic (namely for any realization) construction of the co-algebraic sector underlying the Snyder geometry and (ii) the definition of a non-ambiguous self interacting scalar field theory on this space-time. The first order correction terms of the corresponding Lagrangian are explicitly computed. The possibility to derive Noether charges for the Snyder space-time is also discussed.
\end{abstract}

\pacs{04.60.Bc; 02.40.Gh; 11.10.Nx}

\maketitle 

\section{Introduction}

Snyder space-time has been the first proposal of non-commutative geometry to tame the UV divergences of quantum field theory \cite{Sny,Yan}. The preliminary idea to solve this problem was to use a lattice structure instead of the space-time continuum \cite{Hei}. However, a lattice breaks the Lorentz invariance, posing serious doubts for accepting the theory. A Lorentz invariant discrete space-time has been formulated only by Snyder. The price to pay is a non-commutative structure of space-time. Because of the success of renormalization theory the Snyder program has been abandoned since its rediscovery forty years later by mathematicians \cite{Con,Wor}. Now, the analysis of field theories on non-commutative space-times has become a fundamental area in theoretical physics (for reviews see \cite{DN,Sza}). 

A quantum field theory on the Snyder space-time has however not yet been constructed and thus the removal of divergences by means of non-commutativity effects has not yet been proved. In this paper we construct a self interacting classical scalar field theory on this space-time. This model can be considered as the starting point for the quantum analysis. 

The non-commutativity of the Snyder space-time is encoded in the commutator between the coordinates which is proportional to the (undeformed) Lorentz generators. The Poincar\'e symmetry underlying this space-time is undeformed at the algebraic level, while the co-algebraic sector is (highly) non-trivial. In a previous paper \cite{BM08} we have shown that, by using the concept of realizations, there exists infinitely many deformed Heisenberg algebras all compatible with this geometry. This freedom can be understood as the freedom in choosing momentum coordinates.

We here complete the previous analysis by studying the co-product and star-product structures underlying the model. Equipped by this technology, we construct a scalar field theory on this non-commutative space-time. Our main goal is to define the theory without ambiguities and without needing supplementary structures (as a deformed measure) necessary in extra dimensional approaches. The momentum space of the Snyder-deformed Poincar\'e group has not a Lie group structure since it is given by the coset $SO(4,1)/SO(3,1)$, i.e. the de Sitter space. The co-product and the induced star-product turn out to be {\it non-associative}. This feature represents the main obstacle in studying field theories on the Snyder space-time. Such a kind of deformation of the Poincar\'e group cannot be recovered within the classification \cite{Wor96}, because only deformations preserving the co-associativity are considered. The language of Hopf algebras \cite{qgroup} does not apply straightforwardly to the Snyder space-time geometry.

The non-associativity propriety obstructed the analysis of the Snyder geometry with respect to other non-commutative space-times. For example $\kappa$-Minkowski, a particular case of Lie algebra type space-time, has been developed at different levels specifying star products \cite{Mel53,Fre06}, differential calculus \cite{Sit,MK09}, scalar field theory \cite{ghost,DLW,AAD,Mel80} and conserved charges \cite{AAA,AM1,AM2}. The key difference between Snyder and $\kappa$-Minkowski is that in the latter the momentum space has the structure of a non-abelian Lie group and thus the co-product is non-commutative, but still associative. In particular, the Snyder space-time is not a special case of $\kappa$-Minkowski \cite{Wess1,Wess2}. In fact, as clarified in \cite{Mel09}, $\kappa$-spaces are based on Lie algebra while Snyder space is grounded on trilinear commutations relations. 

Our approach is based on the framework of realizations by which we bypass the non-associativity and clearly define the self interacting theory. The theory we construct lives on the non-commutative space-time and its dual has the momentum space given by a coset. Our analysis deals with the 4d Lorentzian model and no extra dimensional structures are invoked. Moreover, our theory is general as we consider all realizations of the geometry, differently to the previous approaches (for other attempts to define a scalar field theory on Snyder space-time see \cite{GL1,BST,ES,GL2}). The frameworks usually adopted are recovered as particular cases of our construction.   
   
The Snyder space-time is linked to doubly special relativity models \cite{Kov,Guo}, loop quantum gravity \cite{LO} and two-time physics \cite{tt}. In particular in \cite{bat}, we have shown that a Snyder-deformed quantum cosmology predicts a big-bounce phenomenology as in loop quantum cosmology \cite{lqc} (for other comparisons between deformed and loop cosmologies see \cite{BM07,BMtaub,BLM}).

The paper is organized as follows. In Section II we describe the algebraic structure of the Snyder space-time. In Section III the co-algebraic sector underlying the non-commutative geometry is analyzed in detail. Section IV is devoted to the formulation of the scalar field theory on this space-time. Finally, in Section V the first order corrections are computed. Concluding remarks follow. 

We adopt units such that $\hbar=c=1$, the signature given by $\eta_{\mu\nu}=\text{diag}(-,+,...,+)$ and the index convention $\{\mu,\nu,...\}\in\{0,...,n\}$.

\section{Snyder space-time}

In this Section we describe the non-commutative Snyder space-time geometry. We discuss the realizations of such a geometry as well as the dispersion relation underlying the model. 

\subsection{Deformed Heisenberg algebras}

Let us consider a $(n+1)$-dimensional Minkowski space-time such that the commutator between the coordinates has the non-trivial structure 
\be\label{snyalg}
[\tilde x_\mu,\tilde x_\nu]=s M_{\mu\nu}\,,
\ee 
where $\tilde x_\mu$ denote the non-commutative coordinates and $s\in\mathbb R$ is the deformation parameter with dimension of a squared length. We demand that the symmetries of such a space are described by an undeformed Poincar\'e algebra. This means that both Lorentz generators $M_{\mu\nu}=-M_{\nu\mu}=i(x_\mu p_\nu-x_\nu p_\mu)$ and translation generators $p_\mu$ satisfy the standard commutation relations
\bea\label{mmmp}
[M_{\mu\nu},M_{\rho\sigma}]&=&\eta_{\nu\rho}M_{\mu\sigma}-\eta_{\mu\rho}M_{\nu\sigma}-\eta_{\nu\sigma}M_{\mu\rho}+\eta_{\mu\sigma}M_{\nu\rho}\nonumber\\
\left[p_\mu,p_\nu\right]&=&0\,.
\eea
We also assume that momenta and non-commutative coordinates transform as undeformed vectors under the Lorentz algebra, i.e. the commutators
\bea\label{commp}
[M_{\mu\nu},p_\rho]&=&\eta_{\nu\rho}p_\mu-\eta_{\mu\rho}p_\nu\,,\\\label{commx}
[M_{\mu\nu},\tilde x_\rho]&=&\eta_{\nu\rho}\tilde x_\mu-\eta_{\mu\rho}\tilde x_\nu\,,
\eea
hold. The quantity $p^2=\eta^{\mu\nu}p_\mu p_\nu$ is then a Lorentz invariant. 

Relations (\ref{snyalg}-\ref{commx}) define the Snyder space-time
geometry. However, they do not fix the commutator between $\tilde
x_\mu$ and $p_\nu$. In particular, as it was shown in \cite{BM08},
there exists infinitely many possible commutators which are all compatible, in the sense that the algebra closes in virtue of the Jacobi identities, with the above requirements. This feature is understood by means of the concept of realization \cite{JM,MS06,KMS,Mel77,Mel80,Mel53} (for a similar framework see \cite{LG,GP}). A realization on a non-commutative space is defined as a rescaling of the deformed coordinates $\tilde x_\mu$ in terms of ordinary phase space variables ($x_\mu,p_\nu$) as 
\be\label{rea}
\tilde x_\mu=\Phi_{\mu\nu}(p)x_\nu\,.
\ee
The most general $SO(n,1)$-covariant realization of the Snyder geometry reads \cite{BM08}
\be\label{real}
\tilde x_\mu=x_\mu\,\varphi_1(A)+s(xp)p_\mu\,\varphi_2(A),
\ee
in which $\varphi_1$ and $\varphi_2$ are two (dependent) functions of the dimensionless quantity $A=sp^2$ (hereafter the convention $(ab)=\eta^{\mu\nu}a_\mu b_\nu$ is adopted). The function $\varphi_2$ depends on $\varphi_1$ by the relation 
\be\label{varphi}
\varphi_2=\f{1+2\dot\varphi_1\varphi_1}{\varphi_1-2A\dot\varphi_1}\,,
\ee
where dot denotes differentiation with respect to $A$. The generic realization (\ref{real}) is completely specified by the function $\varphi_1$. There are thus infinitely many ways to express, via $\varphi_1$, the non-commutative coordinates (\ref{snyalg}) in terms of the ordinary ones without deforming the original symmetry. The boundary condition $\varphi_1(0)=1$ ensures that the ordinary commutative framework is recovered as soon as $s=0$. The commutator between $\tilde x_\mu$ and $p_\nu$ immediately follows from (\ref{real}) and reads
\be\label{xpcom}
[\tilde x_\mu,p_\nu]=i\left(\eta_{\mu\nu}\varphi_1+s p_\mu p_\nu\varphi_2\right)\,.
\ee
This relation describes a deformed Heisenberg algebra. 

It is also interesting to give the inverse of the realization (\ref{real}), which reads
\be\label{realinv}
x_\mu=\f1{\varphi_1}\left(\tilde x_\mu-\f1{\varphi_1+A\varphi_2}\,s(\tilde xp)p_\mu\varphi_2\right).
\ee
This relation allows us to construct invariants for the
non-commutative framework from those arising in the commutative
one. We only have to demand that the invariants in
($x_\mu,p_\nu$)-coordinates will be sent into the invariants in ($\tilde x_\mu,p_\nu$)-coordinates by means of (\ref{realinv}).  

The use of realizations allow us to give a phase space interpretation of the Snyder space-time. Consider the {\it non-canonical} transformation $x_\mu\rightarrow\Phi_{\mu\nu}(p)x_\nu,\,p_\nu\rightarrow p_\nu$ in an ordinary phase space coordinatized by ($x_\mu,p_\nu$). The Snyder non-commutative geometry results from such a map. This transformation can be a generic function of momenta, but linear in coordinates (for discussions on non-commutative classical mechanics see e.g. \cite{nccm}).  

\subsection{Particular realizations}

The non-commutative Snyder geometry has been analyzed in literature from different points of view \cite{Kov,Guo,LO,tt,GL1,BST,ES,GL2,bat,GB,Gli}, but only two particular realizations of its algebra are usually adopted. These are the Snyder \cite{Sny} and the Maggiore \cite{Mag1,Mag2} types of realizations which are particular cases of (\ref{real}). 

The first realization is the one originally suggested by Snyder. It is recovered from (\ref{real}) if 
\be\label{snyreal}
\varphi_1=1\,,
\ee
which, because of (\ref{varphi}), implies that $\varphi_2=1$. The second realization has been proposed by Maggiore and it appears as soon as 
\be\label{magreal}
\varphi_1=\sqrt{1-sp^2}\,,
\ee
and thus, from (\ref{varphi}), $\varphi_2=0$. The momentum $p_\mu$ is bounded or unbounded depending of the sign of $s$. If $s>0$ the constraint $|p|<1/\sqrt s$ holds. 

Beside these types of realization, the one which realizes the Weyl symmetric ordering is the third interesting one. The Weyl ordering is obtained by the condition
\be
\Phi_{\mu\nu}p_\nu=(\varphi_1\eta_{\mu\nu}+sp_\mu p_\nu\varphi_2)p_\nu=p_\mu,
\ee
which, considering the relation (\ref{varphi}), implies that 
\be\label{weylreal}
\varphi_1=\sqrt{sp^2}\,\cot\sqrt{sp^2}.
\ee
As we said, there are however infinitely many possible realizations of the Snyder space-time geometry. 

\subsection{Dispersion relation}

Let us discuss the fate of the standard dispersion relation $p^2=m^2$ in the Snyder space-time. In particular, we are interested in how different realizations modify this constraint. Consider two momenta $\tilde p_\mu$ and $p_\mu$ in two distinct realizations. Since momenta transform as vectors under the Lorentz symmetry, see (\ref{commp}), the relation
\be\label{tildep}
\tilde p_\mu=p_\mu\,f(A)
\ee
holds. The function $f(A)$, such that $f(0)=1$, depends on the
realization $\varphi_1$ and can be obtained as follows. Let for
example $\tilde p_\mu$ be a momentum in the Maggiore realization (\ref{magreal}), i.e. the commutator $[\tilde x_\mu,\tilde p_\nu]=i\eta_{\mu\nu}\sqrt{1-s\tilde p^2}$ holds. The function $f$ is obtained by inserting (\ref{real}) and (\ref{tildep}) in this relation and reads
\be\label{fa}
f=\f1{\sqrt{\varphi_1^2+A}}\,.
\ee
Notice that $f=1$ as the realization (\ref{magreal}) is taken into account and that, because of (\ref{snyreal}), the Maggiore momentum $p^M_\mu$ is related to the Snyder one $p^S_\mu$ by
\be
p^S_\mu=p^M_\mu\,\sqrt{1+s(p^S)^2}\,.
\ee

As expected from (\ref{commp}), the dispersion relation for $\tilde p^2$ is undeformed, but an effective mass $m_e=m_e(m)$ has to be taken into account. From (\ref{tildep}) and (\ref{fa}), the Snyder-dispersion relation reads
\be
\tilde p^2=\f{m^2}{\varphi_1^2(sm^2)+sm^2}\equiv m^2_e\,.
\ee 
Let us discuss such a formula in the Snyder realization ($\varphi_1=1$). In the low-deformed case ($m^2\ll1/s$) the effective mass is given by $m^2_e\simeq m^2(1-sm^2)$. On the other hand, in the ultra-deformed case ($m^2\gg1/s$ with $s>0$) we have $m^2_e\simeq1/s$ (for $s<0$, $m$ is bounded as $m^2<1/|s|$). 

\section{Co-algebraic sector}

The co-algebraic sector of the Snyder geometry is here analyzed. We firstly focus on a generic framework, i.e. by considering the arbitrary realization (\ref{real}). The two particular realizations (\ref{snyreal}) and (\ref{magreal}) are investigated below. A discussion about the non-associativity follows.

\subsection{General framework}

Deformations of symmetries underlying Snyder space-time (\ref{snyalg})
are contained in the co-algebraic sector of a (non-trivial) quantum
group. Generators $(\tilde x_\mu,p_\mu,M_{\mu\nu})$ form an algebra defined by the commutators (\ref{snyalg})-(\ref{commx}) and (\ref{xpcom}). This is not a Hopf algebra. However, $(p_\mu,M_{\mu\nu})$ generate the Snyder-deformed Poincar\'e group $\mathcal P_S$ whose algebra is a generalization of the Hopf algebra.  

As understood from commutators (\ref{mmmp}) and (\ref{commp}),
the Snyder algebraic sector is the one of an undeformed Poincar\'e
algebra. On the other hand, the co-algebraic sector, defined
by the action of Poincar\'e generators on the Snyder coordinates $\tilde x_\mu$, is deformed. The action of Lorentz
generators is still the standard one because of (\ref{commx}), but the
action of momenta is modified as in (\ref{xpcom}). The Leibniz rule is thus deformed and depends on realizations. As we will see, the co-product for momenta is no longer commutative and neither associative. 

The co-product and star-product structures can obtained from
realizations as follows. Let $\mathbb I$ be the unit element of the space of commutative functions $\psi(x)$. By means of (\ref{real}) the action of a non-commutative function $\tilde\psi(\tilde x)$ on $\mathbb I$ gives \cite{Mel09,Mel12}
\be\label{pro}
\tilde\psi(\tilde x)\triangleright\mathbb I=\psi'(x)\,.
\ee 
This relation provides a map from the non-commutative space of functions to the commutative one. Notice that the commutative function $\psi'(x)$ will be in general different from $\psi(x)$. Consider now a non-commutative plane wave $e^{i(k\tilde x)}$, in which $\tilde x_\mu$ refers to a given realization (\ref{real}) and $k_\mu$ are the eigenvalues of $p_\mu=-i\p/\p x^\mu$. It is then possible to show that \cite{Mel09,Mel12}
\be\label{plawave}
e^{i(k\tilde x)}\triangleright\mathbb I=e^{i(K x)},
\ee
where $K_\mu=K_\mu(k)$ is a deformed momentum (defined below) depending on realizations. The commutative limit $s\rightarrow0$ leads to the standard framework in which $K_\mu=k_\mu$. Consequently, given the inverse transformation $K_\mu^{-1}=K_\mu^{-1}(k)$, we have
\be\label{invk}
e^{i(K^{-1}\tilde x)}\triangleright\mathbb I=e^{i(k x)}.  
\ee
It is worth noting that, in the Weyl realization (\ref{weylreal}), we have $e^{i(k\tilde x)}\triangleright\mathbb I=e^{i(k x)}$ and plane waves are undeformed. 

Let us now consider two plane waves labeled by momenta $k_\mu$ and $q_\mu$, respectively. Their action on the unit element $\mathbb I$ gives 
\be\label{twopla}
e^{i(k\tilde x)}\left(e^{i(q x)}\right)=e^{i(F(k,q)x)}\,. 
\ee
The deformed momentum $K_\mu$ is thus determined by $K_\mu=F_\mu(k,0)$, where the function $F_\mu(k,q)$ specifies the co-product as well as the star-product. It can be obtained by a straightforward implementation of the Campbell-Baker-Hausdorff formula or by the more elegant method developed in \cite{Mel09,Mel12}. 

The star-product between two plane waves is defined by $F_\mu(k,q)$ as
\begin{multline}\label{starpro}
e^{i(k x)}\star e^{i(q x)}\equiv e^{i(K^{-1}(k)\tilde x)}e^{i(K^{-1}(q)\tilde x)}\triangleright\mathbb I=\\
=e^{i(K^{-1}(k)\tilde x)}\left(e^{i(q x)}\right)=e^{i(\mathcal D(k,q) x)}
\end{multline}
in which 
\be\label{copro}
\mathcal D_\mu(k,q)=F_\mu\left(K^{-1}(k),q\right)\,.
\ee
The star-product defines, by means of (\ref{plawave}) and (\ref{invk}), a Weyl mapping from the commutative to the non-commutative spaces provided by a one-to-one correspondence between $e^{i(k\tilde x)}$ and $e^{i(K x)}$. The co-product for momenta $\Delta p_\mu$ (and the corresponding Leibniz rule) is obtained from $\mathcal D_\mu(k,q)$ as 
\be\label{coprotot}
\Delta p_\mu=\mathcal D_\mu(p\otimes1,1\otimes p)\,.
\ee
In particular, the function $\mathcal D_\mu$ describes the non-abelian sum of momenta in the Snyder non-commutative space-time, i.e.
\be\label{dmu}
\mathcal D_\mu(k,q)=k_\mu\oplus q_\mu\neq k_\mu+q_\mu\,.
\ee
As soon as the non-commutativity effects (in our case the parameter $s$) are switched off, the ordinary abelian rule $\mathcal D_\mu(k,q)=k_\mu+q_\mu$ is recovered. By means of (\ref{starpro}), it is possible to obtain the star-product between two generic functions $f$ and $g$ of commuting coordinates (see for example \cite{MS06,Mel53,Mel12}). Adopting the plane waves relation (\ref{starpro}), the general result for the star product stands as
\be\label{starprofg}
(f\star g)(x)=\lim_{\substack{y\rightarrow x\\z\rightarrow x}}e^{ix_\mu\left(\mathcal D^\mu(p_y,p_z)-p^\mu_y-p^\mu_z\right)}f(y)g(z)\,.
\ee
Star product is a binary operation acting on the algebra of functions defined on the ordinary commutative space and it encodes features reflecting the non-commutative nature of Snyder space-time (\ref{snyalg}). The star product is uniquely defined, but its concrete form is related to a particular realization and vice versa. For any realization the star product (\ref{starpro}), and then (\ref{starprofg}), is non-associative. The corresponding co-product (\ref{coprotot}) is non-co-associative. Such a result has been confirmed by the recent analysis \cite{GL2} also.   

This construction is well defined and allows us to obtain, from
realizations (\ref{real}), both co-product and star-product structures
underlying the Snyder space-time. The inverse path is also meaningful:
starting from a star-product (or a co-product) it is always possible
to recover an information about realization we are working in. However, as we shall see, to construct a scalar field theory on Snyder space-time it is more suitable to deal with realizations instead of star-products. The non-associativity of the star-product in fact poses severe challenges in defining interaction terms.

Let us now compute the co-product $\Delta p_\mu$, at the first order in $s$, for a generic realization (\ref{real}). Expanding the realization function $\varphi_1$ as $\varphi_1=1+c_1sp^2+\mathcal O(s^2)$ and considering (\ref{fa}), we obtain
\be\label{gencopro}
\Delta p_\mu=\Delta_0 p_\mu+s\,\Delta_1 p_\mu+\mathcal O(s^2) 
\ee
\be\nonumber
\Delta_0 p_\mu=p_\mu\otimes1+1\otimes p_\mu
\ee
\begin{multline}\nonumber
\Delta_1 p_\mu=\left(c-\f12\right)p_\mu\otimes p^2+\left(2c-\f12\right)p_\mu p_\nu\otimes p^\nu+\\
+c\left(p^2\otimes p_\mu+2p_\nu\otimes p^\nu p_\mu\right)
\end{multline} 
where $c=(2c_1+1)/2$. Here $\Delta_0 p_\mu$ and $\Delta_1 p_\mu$ denote the co-product at the zero and first order is $s$, respectively. The Maggiore type of realization (\ref{magreal}) is defined by $c_1=-1/2$ and thus it is recovered as $c=0$. The Snyder one (\ref{snyreal}) appears for $c_1=0$ and thus $c=1/2$, while the Weyl one (\ref{weylreal}) for $c=1/6$. The co-product (\ref{gencopro}) defines the Snyder non-abelian sum in a generic realization. The star-product is obtained from (\ref{starpro}).

To complete the analysis of the co-algebraic sector we need to specify the co-product $\Delta M_{\mu\nu}$ of the Lorentz generators, as well as the antipode $S(g)$ and the co-unit $\varepsilon(g)$ for any element $g$ of $\mathcal P_S$. Because of relations (\ref{mmmp})-(\ref{commx}), the co-product $\Delta M_{\mu\nu}$ is trivial, i.e.
\be\label{copromom}
\Delta M_{\mu\nu}=M_{\mu\nu}\otimes1+1\otimes M_{\mu\nu}\,.
\ee
The antipode $S(g)$ is defined by the equation
\be
\mathcal D\left(g,S(g)\right)=g\oplus S(g)=0\,.
\ee
From (\ref{gencopro}) and (\ref{copromom}), we immediately realize that the antipode is not deformed for any $g=(p_\mu,M_{\mu\nu})$, that is
\be\label{anti}
S(p_\mu)=-p_\mu\, \qquad S(M_{\mu\nu})=-M_{\mu\nu}\,.
\ee
Because different momenta are related by (\ref{tildep}), the antipode $S(p_\mu)$ is exactly (not only at the first order) trivial in all realizations. On the other hand, the co-unit $\varepsilon(g)$ is also trivial for any $g\in\mathcal P_S$. Finally, we observe that co-product for momenta (\ref{coprotot}) is covariant becouse of (\ref{copromom}), i.e. the relation
\be
[\Delta M_{\mu\nu},\Delta p_\rho]=\eta_{\nu\rho}\Delta p_\mu-\eta_{\mu\rho}\Delta p_\nu
\ee
holds. This expression is the co-algebraic counter term of (\ref{commp}).

Summarizing, the Snyder-deformed Poincar\'e group $\mathcal P_S$ is characterized as follows. The Lorentz symmetry is undeformed at both algebraic and co-algebraic level. The deformations are encoded in the co-product (\ref{coprotot}) only, which in particular is non-co-associative. The corresponding star-product (\ref{starpro}) is non-associative and a homomorphism relates these structures. The algebraic sector is then compatible with the co-algebraic one. Therefore, the generators $(p_\mu,M_{\mu\nu})$ of $\mathcal P_S$ form a generalized Hopf algebra, which we shall denote as a {\it non-co-associative Hopf algebra}.  

\subsection{Particular realizations}

We now study the co-product structure underlying the two particular realizations (\ref{snyreal}) and (\ref{magreal}). In both cases a closed form of the co-product arises.

Let us firstly consider the Maggiore realization (\ref{magreal}). The basic function $F_\mu(k,q)$ in (\ref{twopla}) is given by
\be\label{fmag}
F_\mu=q_\mu+k_\mu\sqrt{1-A_q}\,\f{\sin\sqrt{A_k}}{\sqrt{A_k}}-sk_\mu(k q)\,\f{1-\cos\sqrt{A_k}}{\sqrt{A_k}}
\ee
where $A_p=sp^2$. The ordinary function $F_\mu=k_\mu+q_\mu$ is recovered in the $s=0$ case. From (\ref{fmag}) one immediately obtains the deformed momentum $K_\mu(k)$ which reads
\be
K_\mu=F_\mu(k,0)=k_\mu\f{\sin\sqrt{A_k}}{\sqrt{A_k}}\,. 
\ee
The co-product $\Delta p_\mu$ follows from (\ref{copro}) and it is given, in terms of the realization function $\varphi_1=\sqrt{1-sp^2}$, by
\be
\Delta p_\mu=p_\mu\otimes\varphi_1-\f s{1+\varphi_1}\,p_\mu p_\nu\otimes p^\nu+1\otimes p_\mu\,.
\ee
Such a co-product (namely the addition rule (\ref{dmu})) is non-co-associative. The order by which we sum the momenta becomes important. As $s=0$ we have $\varphi_1=1$ and the trivial co-product $\Delta_0 p_\mu=p_\mu\otimes1+1\otimes p_\mu$ is recovered. The first-order term coincides with (\ref{gencopro}) for $c=0$.

Let us now analyze the Snyder realization (\ref{snyreal}). In this case the function $F_\mu(k,q)$ in (\ref{twopla}) reads
\bea\label{fsny}
F_\mu&=&g\,(h\,k_\mu+q_\mu)\\\nonumber
g&=&\left(\cos\sqrt{A_k}-\sqrt{\f s{k^2}}(k q)\sin\sqrt{A_k}\right)^{-1}\\\nonumber
h&=&\f1{k^2}\left(\sqrt{\f{k^2}s}\sin\sqrt{A_k}+(k q)(\cos\sqrt{A_k}-1)\right)\,.
\eea
The deformed momentum $K_\mu(k)$ is then given by 
\be
K_\mu=F_\mu(k,0)=k_\mu\f{\tan\sqrt{A_k}}{\sqrt{A_k}}\,.
\ee
The ordinary framework is restored as $s=0$. The co-product directly follows from (\ref{copro}) and reads
\begin{multline}
\Delta p_\mu=\f1{1-sp_\nu\otimes p^\nu}\left(\f{}{}p_\mu\otimes1+\right.\\
-\left.\f s{1+\sqrt{1+A_p}}\,p_\mu p_\nu\otimes p^\nu+\sqrt{1+A_p}\otimes p_\mu\right).
\end{multline}
Also in this case the co-product is non-co-associative. The first order-term coincides with the $c=1/2$ case of (\ref{gencopro}).

\subsection{On the non-associativity}

The relation between special relativity and the Snyder geometry allows us to better understand the physical meaning of the non-associativity. 

Special relativity can be analyzed (and derived) from a non-commutative point of view \cite{girliv}. Consider the Galileo group $ISO(3)=SO(3)\cdot\mathbb R^3$. Speeds generate translations and the speed space $\mathbb R^3$ can be identified as $\mathbb R^3\sim ISO(3)/SO(3)$. A manifold of this type is a coset space. In this case it has the (Lie) group structure. Special relativity can be viewed as arising from the deformation of $\mathbb R^3$ into the curved space $\mathcal C=SO(3,1)/SO(3)$. This operation sends the Galileo group into the Lorentz one $SO(3,1)=SO(3)\cdot \mathcal C$ (this is the Cartan decomposition of the Lorentz group \cite{Lie}). The coset $SO(3,1)/SO(3)$ is nothing but the (hyperbolic) boosts space, but it is not a Lie group. In fact the product between two boosts is not longer a boost, but an element of the full Lorentz group $SO(3,1)$. The composition of speeds can be extracted from a co-product structure. It turns out that the composition of (non-collinear) speeds is no longer commutative and neither associative. A physical manifestation of non-associativity is the well-known Thomas precession \cite{thomas}. From a mathematical point of view, the non-associativity is a consequence of the fact that the coset space is not a group manifold.

The Snyder space-time geometry can be viewed from the same perspective. Consider the Poincar\'e group $\mathcal P=SO(3,1)\cdot\mathbb R^4$. As above, the momentum space $\mathbb R^4$ can be viewed as the coset $\mathbb R^4\sim\mathcal P/SO(3,1)$ and of course it is a group manifold. Deforming the momentum space into the de Sitter space $d\mathcal S=SO(4,1)/SO(3,1)$ we recover the Snyder non-commutative geometry. This is the original formulation made by Snyder himself \cite{Sny}. The Snyder-deformed Poincar\'e group $\mathcal P_S$ is then factorized as 
\be\label{pdef}
\mathcal P_S=SO(3,1)\cdot d\mathcal S,
\ee
showing that the Lorentz symmetry is undeformed. On the other hand, the translation sector of this (quantum) group is deformed consistently to (\ref{snyalg}). As in the previous case, the coset $d\mathcal S$ is not a Lie group. The non-co-associativity of Snyder co-product can be traced back to this feature. 

\section{Scalar field theory}

In this Section we construct the scalar field theory on the 4d Snyder non-commutative space-time. We first consider the Fourier transformation and define the Snyder scalar field and then we write down the action for the theory. A comparison with other approaches follows.

\subsection{Preliminaries}

We define a scalar field $\tilde\phi(\tilde x)$ on the Snyder non-commutative space-time by means of the Fourier transformation as 
\be\label{ncsca}
\tilde\phi(\tilde x)=\int [dk]\,\hat\phi(k)\,e^{i(K^{-1} \tilde x)}\,.
\ee
The integration measure $[dk]$ is a priori deformed depending on the antipode $S(k_\mu)$. However, as we have previously seen, it is trivial in any realizations. The measure in (\ref{ncsca}) is thus the ordinary one
\be
[dk]=\f{d^4k}{(2\pi)^4}\,.
\ee 
Let us now consider the action of Snyder scalar field (\ref{ncsca}) on the identity $\mathbb I$. By means of (\ref{invk}), this operation gives
\be
\tilde\phi(\tilde x)\triangleright\mathbb I=\phi(x)\,,
\ee
which ensures the Lorentz scalar behavior of the model. As a further step we consider the quadratic term $\tilde\phi^2(\tilde x)\triangleright\mathbb I$. Given the definition (\ref{ncsca}) and remembering (\ref{twopla}), (\ref{starpro}) and (\ref{copro}), we obtain
\be\label{phi2}
\tilde\phi^2(\tilde x)\triangleright\mathbb I=\left(\phi\star\phi\right)(x)\,.
\ee
We have thus recovered the star-product structure. 

Let us now discuss the notion of a real and complex Snyder scalar
field. Firstly, because of triviality of the antipode, the conjugation
is also an ordinary one. Secondly, the non-commutative coordinates $\tilde x_\mu$ have to be hermitian operators in any given realization. All the commutators given above are invariant under the formal anti-linear involution ``$\dag$''
\be
\tilde x_\mu^\dag=\tilde x_\mu, \qquad p_\mu^\dag=p_\mu, \qquad M_{\mu\nu}^\dag=-M_{\mu\nu}\,,
\ee
where the order of elements is inverted under the involution. On the other hand, the realization (\ref{real}) is in general not hermitian. The hermiticity condition can be immediately implemented as soon as the expression
\be
\tilde x_\mu=\f12\left(x_\mu\varphi_1+s(x\,p)p_\mu\varphi_2+\varphi_1^\dag x_\mu^\dag+s\,\varphi_2^\dag p_\mu^\dag(x\,p)^\dag\right)
\ee 
is taken into account. However, the physical results do not depend on
the choice of the representation as long as there exists a smooth limit $\tilde x_\mu\rightarrow x_\mu$ as $s\rightarrow0$. We can thus restrict our attention to non-hermitian realization only. Consequently, we focus on the real Snyder scalar field theory, while the complex one can be straightforwardly defined.

\subsection{Action for scalar field theory}

We are now able to construct a Lagrangian for the non-commutative
scalar field (\ref{ncsca}). Let us start by analyzing how the ordinary
kinematic term $(\p_\mu\phi)(\p^\mu\phi)$ is changed in the Snyder
space-time. Following the previous reasonings, the corresponding term
in the non-commutative framework is given by
$(\p_\mu\tilde\phi)(\p^\mu\tilde\phi)\triangleright\mathbb I$ (notice
that the derivative is still with respect to the commutative
coordinates, i.e. $\p_\mu=\p/\p x^\mu$). Such a term, expressed by
means of the Fourier transformation (\ref{ncsca}), is uniquely
defined. In fact, in order that the differentiation makes sense, we have firstly to project the plane waves on $\mathbb I$ and then act on these by differentiation. By using (\ref{twopla}) and (\ref{anti}), the relation
\be\label{relatio}
\left(\p_\mu e^{i(K^{-1}\tilde x)}\right)e^{i(qx)}=i(\mathcal D_\mu-q_\mu)e^{i(\mathcal D(k,q)x)}=ik_\mu e^{i(\mathcal D(k,q)x)}\,
\ee
follows. The kinematic part, considering (\ref{invk}) and (\ref{relatio}), is then given by
\begin{widetext}
\be\label{dmuphi2}
(\p_\mu\tilde\phi)(\p^\mu\tilde\phi)\triangleright\mathbb I=\int[d^2k]\,\hat\phi_{k_1}\hat\phi_{k_2}\left(\p_\mu e^{i(K_1^{-1}\tilde x)}\right)\p^\mu\left(e^{i(K_2^{-1}\tilde x)}\triangleright\mathbb I\right)=-\int[d^2k]\,\hat\phi_{k_1}\hat\phi_{k_2}(k_1k_2)\,e^{i(\mathcal D(k_1,k_2)x)}\,,
\ee
\end{widetext}
where $[d^nk]=[dk_1]...[dk_n]$ and $\phi_k=\phi(k)$. This expression leads to the correct ordinary result as $s=0$. As in (\ref{phi2}), the star-product prescription leads to the same result with respect to our construction:
\be
(\p_\mu\tilde\phi)(\p^\mu\tilde\phi)\triangleright\mathbb I=(\p_\mu\phi)\star(\p^\mu\phi)\,.
\ee
The action for a non-interacting massive scalar field on Snyder space-time then reads
\begin{multline} 
I=\int d^4x\left(\p_\mu\tilde\phi\,\p^\mu\tilde\phi+m^2\,\tilde\phi^2\right)\triangleright\mathbb I=\\
=\int d^4x\left[(\p_\mu\phi)\star(\p^\mu\phi)+m^2(\phi\star\phi)\right]\,.
\end{multline}
Because of the antipode (\ref{anti}), the action in the momentum space can be trivially written. 

The non-commutativity effects are thus summarized within the co-product (\ref{gencopro}), i.e. within the non-abelian sum $\mathcal D_\mu(k_1,k_2)$. The non-commutative corrections to the ordinary theory depend on realizations. For each type of realization different actions appear.

Finally, we investigate the role of self interactions. In particular, we consider the cubic $\tilde\phi^3(\tilde x)\triangleright\mathbb I$ and quartic $\tilde\phi^4(\tilde x)\triangleright\mathbb I$ interaction terms. These terms can be immediately obtained. The generalization of (\ref{twopla}) to three plane waves, considering also (\ref{starpro}), reads
\be
e^{i(K_3^{-1}\tilde x)}\left(e^{i(K_2^{-1}\tilde x)}\left(e^{i(K_1^{-1}\tilde x)}\triangleright\mathbb I\right)\right)=e^{i(\mathcal D_3(k_3,k_2,k_1) x)}\,,
\ee
in which $(\mathcal D_3)_\mu(k_3,k_2,k_1)=\mathcal D_\mu(k_3,\mathcal D(k_2,k_1))$. This defines the cubic term 
\be
\tilde\phi^3(\tilde x)\triangleright\mathbb I=\left(\phi\star(\phi\star\phi)\right)(x).
\ee
The quartic term $\tilde\phi^4(\tilde x)\triangleright\mathbb I$ is determined in the same way. Given four plane waves we have
\be\label{phi4}
e^{i(K_4^{-1}\tilde x)}\left(e^{i(K_3^{-1}\tilde x)}\left(e^{i(K_2^{-1}\tilde x)}\left(e^{i(K_1^{-1}\tilde x)}\triangleright\mathbb I\right)\right)\right)=e^{i(\mathcal D_4 x)}\,,
\ee
and therefore
\be
\tilde\phi^4(\tilde x)\triangleright\mathbb I=\left(\phi\star(\phi\star(\phi\star\phi))\right)(x)\,,
\ee
where $(\mathcal D_4)_\mu=\mathcal D_\mu(k_4,\mathcal D_3(k_3,k_2,k_1))$.  

Summarizing, we have defined a Lagrangian density for a self interacting scalar field on the Snyder non-commutative space-time geometry. Our framework, which is based on realizations, uniquely fixes the theory. This is relevant because the co-product is non-co-associative (the corresponding star-product is non-associative). This feature would lead, a priori, to a non-unique definition of the model. Such a shortcoming is bypassed in our construction.

\subsection{Relation with other approaches}

Our construction differs with respect to the usual ones in two main points: the dimensions of the structure underlying the theory and the adopted algebra. 

The scalar field theory on Snyder space-time is usually
formulated by considering a five dimensional structure \cite{GL1,BST,ES,GL2}. The same
happens for the field theories on $\kappa$-Minkowski \cite{ghost,DLW,AAD}. In particular, the momentum space is the de Sitter section in a five dimensional flat space and a deformed Fourier measure is thus needed to ensure the Lorentz invariance \cite{GL1,BST,ES,GL2}. In $\kappa$-Minkowski, a five dimensional differential structure predicts some unphysical ghost modes \cite{ghost} (to overcame this feature a twist deformation of the symmetry has been proposed \cite{twist1,twist2}). On the other hand, our theory is defined on a four dimensional space-time. No extra measures are needed and the theory has the same field structure of the commutative framework. The Snyder deformed symmetry algebra is the original undeformed one and only the co-product structure changes. Interesting, this is exactly the framework arising from the twist formulation of non-commutative field theories \cite{twist1,twist2,Mel80}.  

The second difference with respect to other approaches is that our theory is generic. All the possible realizations of the algebra are taken into account. The other attempts to construct a scalar field theory on the Snyder space-time are in fact based on a particular realization only. Our theory in the Snyder type of realization (\ref{snyreal}) corresponds, up to the momentum-space duality, to the previous proposals \cite{GL1,BST,ES,GL2}.  
  
\section{First order corrections}

In this Section we explicitly compute the generic non-commutative corrections, up to the first order in $s$, to the commutative theory. 

As we have seen, all the non-commutative informations are summarized in the non-abelian sum (\ref{dmu}), namely in the co-product (\ref{gencopro}). We are thus interested in the function $(\mathcal D_4)_\mu=(\mathcal D_4)_\mu(k_4,k_3,k_2,k_1)$ defined in (\ref{phi4}). The functions $(\mathcal D_3)_\mu(k_3,k_2,k_1)$ and $\mathcal D_\mu(k_2,k_1)$, which define the cubic and quadratic terms, are clearly recovered from this one as soon as $k_4=0$ and $k_4=k_3=0$, respectively. The $(\mathcal D_4)_\mu$ function can be expanded in the deformation parameter $s$ as
\be\label{expg}
(\mathcal D_4)_\mu=(\mathcal D_4)^0_\mu+s(\mathcal D_4)^1_\mu+\mathcal O(s^2)
\ee
\be\nonumber
(\mathcal D_4)^0_\mu=(k_1)_\mu+(k_2)_\mu+(k_3)_\mu+(k_4)_\mu
\ee
\be\nonumber
(\mathcal D_4)^1_\mu=\alpha(k_1)_\mu+\beta(k_2)_\mu+\gamma(k_3)_\mu+\delta(k_4)_\mu\,,
\ee
where the superscript denotes the order in $s$. 

The correction term $(\mathcal D_4)^1_\mu$ depends on realizations through $\alpha=\alpha(\varphi_1)$, $\beta=\beta(\varphi_1)$, $\gamma=\gamma(\varphi_1)$ and $\delta=\delta(\varphi_1)$. These functions are given by
\begin{multline}
\alpha=c\left[k_2^2+k_3^2+k_4^2+2(k_1k_2+k_3k_2+k_3k_1+\right.\\
\left.+k_4k_3+k_4k_2+k_4k_1)\right]\,,
\end{multline}
\begin{multline}
\beta=\left(c-\f12\right)k_1^2+\left(2c-\f12\right)(k_1k_2)+c\,[k_3^2+k_4^2+\\
+2(k_3k_2+k_3k_1+k_4k_3+k_4k_2+k_4k_1)]\,,
\end{multline}
\begin{multline}
\gamma=\left(c-\f12\right)(k_1+k_2)^2+\left(2c-\f12\right)(k_3k_2+k_3k_1)+\\
+c\,[k_4^2+2(k_4k_3+k_4k_2+k_4k_1)]\,,
\end{multline}
\be
\delta=\left(c-\f12\right)(k_1+k_2+k_3)^2+\left(2c-\f12\right)(k_4k_3+k_4k_2+k_4k_1).
\ee
The value of the constant $c$ determines the realization in which we are working. The Snyder (\ref{snyreal}), the Maggiore (\ref{magreal}) and the Weyl (\ref{weylreal}) types of realization are respectively recovered for $c=1/2,0,1/6$.

\section{Concluding remarks}

In this paper we have constructed a scalar field theory on the Snyder non-commutative space-time. The next step will be the quantization of the model in order to investigate the fate of UV divergences and thus fully analyze the Snyder proposal.

We have shown that the deformations of symmetries are all contained in the co-algebraic sector and that the co-product is non-co-associative. By using the realizations of the Snyder algebra we have constructed a well defined (namely non-ambiguous) self interacting scalar field theory. The ambiguities carried out by the non-associative sum of momenta (and thus the non-associative star-product) have been overcome by the use of realizations. By means of a map between the non-commutative functions and the commutative ones, a scalar field action has been constructed. This theory has been directly defined on the space-time and, since the Fourier space has been identified with the de Sitter space, it is dual to a field theory over the coset $SO(4,1)/SO(3,1)$. Finally, we have computed the first order corrections in a generic realization.

As last point, it is interesting to mention that we can construct Noether charges for the Snyder space-time. As was shown in \cite{AM1,AM2}, the key ingredient to build Noether charges in a non-commutative theory is a Poisson map between the deformed and the undeformed spaces of solutions of the Klein-Gordon equation. In our framework this kind of map is given by the projection of non-commutative functions on the ``vacuum'', as in (\ref{pro}). By using this map it is possible to induce a symplectic structure on the space of the non-commutative functions and thus obtain a conserved symplectic product defining charges. This analysis will be reported elsewhere \cite{BatMel10}.  

\section*{Acknowledgments}

We would like to thank Andjelo Samsarov for useful discussions and for a critical reading of the manuscript. Antonino Marcian\`o is thanked for comments. Part of this work is supported by the Ministry of Science and Technology of the Republic of Croatia under contract No. 098-0000000-2865.

\end{document}